# A Perceptual Based Motion Compensation Technique for Video Coding

Amin Banitalebi, Said Nader-Esfahani, Alireza Nasiri Avanaki
School of Electrical and Computer Engineering
University of Tehran, Tehran 14395-515, Iran
banitalebi@ieee.org

*Abstract*—Motion estimation is one of the important procedures in the all video encoders. Most of the complexity of the video coder depends on the complexity of the motion estimation step. The original motion estimation algorithm has a remarkable complexity and therefore many improvements were proposed to enhance the crude version of the motion estimation. The basic idea of many of these works were to optimize some distortion function for mean squared error (MSE) or sum of absolute difference (SAD) in block matching But it is shown that these metrics do not conclude the quality as it is, on the other hand, they are not compatible with the human visual system (HVS). In this paper we explored the usage of the image quality metrics in the video coding and more specific in the motion estimation. We have utilized the perceptual image quality metrics instead of MSE or SAD in the block based motion estimation. Three different metrics have used: structural similarity or SSIM, complex wavelet structural similarity or CW-SSIM, visual information fidelity or VIF. Experimental results showed that usage of the quality criterions can improve the compression rate while the quality remains fix and thus better quality in coded video at the same bit budget.

I. INTRODUCTION

In the few past years, the video compression and coding paradigm has been under a lot of attention and the subject of extensive research. Applications such as video broadcasting, video transmission over the networks and video streaming on the internet as well as the movie industry, explored the necessity of the compression of the video signals further and further.

In the original H.261 video coding protocol the main part of the encoder is motion estimation stage. A simple block matching is performed on the current frame (or reference image) and the next frame (or target image). SAD is used as the distortion measure for block matching i. e. the block from the next frame with the smallest SAD between the reference block is selected as the best match. There is usually a search range for the blocks to be compared to the reference block. In the simplest case the search range can be a block with side equal to 3 times bigger than the reference block [1], [2], [3]. Figure 1 shows a typical diagram of the motion estimation.

After the introduction of H.261, researchers had consumed much effort to improve the performance of H.261 video coder. Several improvements were added to the crude version of the motion estimation for improving the compression ratio and reducing the computational complexity [3]-[11]. H.263 added more features such as supporting more picture formats, different GOB (group of blocks) structure, unrestricted motion vectors, *P* and *B* frames mode, advanced prediction mode and also half-pel motion estimation which allows the resolution of motion vectors to be half pixel [6]. There is a brief comparison between H.261 and H.263 in [3]. H.263+ and H.263++ were introduced after H.263 and H.264 and MPEG4 are the last version of the coders with many optional modes [7].

Block matching motion estimation algorithms such as the three step search [10], [12] and the diamond search [13], [14] algorithms are currently being used in video coding schemes as alternatives to full search algorithms. Many improvements to reduce the complexity of these methods proposed. In the almost all of the mentioned video coding standards, there is a lack of consideration of the human perception system in the motion estimation procedure. Using a SAD or MSE criterion would perhaps lead to less computational complexity but it is not optimal in case of coding rate or quality of the compressed video. In the motion compensation of H.264 the best matching blocks and the best prediction modes are chosen by Lagrange cost function whose distortion function is SAD.

It has been shown that MSE is not a reliable quality criterion when we are dealing with perceptually important signals such as images and videos [15]-[18]. A. C. Bovic and his team proposed new quality metrics for images and videos such as SSIM, CW-SSIM and VIF that are based on human visual system. An overview of image quality assessment (IQA) can be found in [16].

In this paper, we utilized the usage of the image quality metrics instead of SAD or MSE in the motion estimation and compensation of the video. We used SSIM [17], CW-SSIM [18] and VIF [15] as quality criterion in motion estimation and explored the effect and trade-offs of utilization of these metrics.

The rest of this paper is organized as follows: section II is a brief introduction to image quality metrics commonly used in quality assessment context. Section III contains the specifications of the dataset. Section IV is dedicated to explaining our method and section V describes the simulation results while section VI is conclusion and section VII contains the references.

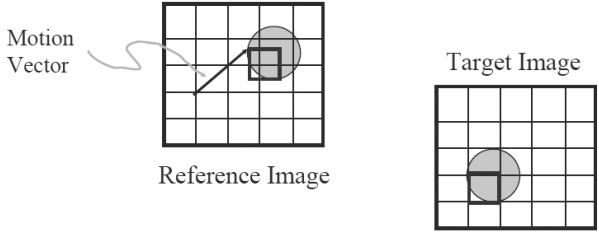

Figure 1. Motion vector & motion compensation

## II. UTILIZED PERCEPTUAL QUALITY METRIC

As mentioned before, it is proven that without modeling the human vision system one cannot introduce an acceptable quality criterion for images and this was the key idea to the generation of HVS based quality metrics. In this section we briefly introduce main criterions for calculate the quality of images and videos.

### A. Sum of Absolute Difference

The most common block distortion measure (BDM) is sum of absolute difference which gives almost similar performance of using MSE but with much less computations. Suppose the block size is $N \times N$ and we want to calculate SAD of a block located at $(x,y)$ in the current frame $F_t$ matched against a block with a displacement $(u,v)$ from $(x,y)$ in previous frame $F_{t-1}$, then SAD is simply defined as:

$$SAD_{x,y}(u,v) = \sum_{i=0}^{N-1}\sum_{j=0}^{N-1} |F_t(x+i, y+j) - F_{t-1}(x+i+u, y+j+v)| \quad (1)$$

SAD is used in motion compensation mainly because of its simplicity and little computational cost implied by SAD to the whole encoding procedure.

### B. Mean Squared Error

Due to its simplicity, MSE (mean square error) has been the dominant quantitative performance metric in the field of signal processing for many years. But it is shown that there is a big lack of accuracy in MSE when dealing with perceptually important signals such as speech, images and video signals. For those applications newly developed perceptual quality metrics are used.

The MSE and PSNR between two 8-bit image signals **x**, and **y** are formulated as:

$$MSE = \frac{1}{N}\sum_{i=1}^{N}(x_i - y_i)^2 \quad (2)$$

$$PSNR = 10\log\frac{255^2}{MSE} \quad (3)$$

Where $x_i, y_i$ are the gray level intensity values of the pixels.

### C. Structural Similarity

SSIM (structural similarity) is a popular and state of the art quality metric in the field of image processing. We mainly use the primitive formulation of the Wang and Bovic [16], [17] in which, the target application is quality assessment of the images. SSIM is a function of luminance, contrast and a local structure function of images **x** and **y**. Local structural similarity is usually formulated as:

$$S(x,y) = l(x,y).c(x,y).s(x,y) = \left(\frac{2\mu_x\mu_y + C_1}{\mu_x^2 + \mu_y^2 + C_1}\right).\left(\frac{2\sigma_x\sigma_y + C_2}{\sigma_x^2 + \sigma_y^2 + C_2}\right).\left(\frac{\sigma_{xy} + C_3}{\sigma_x\sigma_y + C_3}\right) \quad (4)$$

Where $\mu_x$ and $\mu_y$ are (respectively) the local sample means of **x** and **y**, $\sigma_x$ and $\sigma_y$ are (respectively) the local sample standard deviations of **x** and **y**, and $\sigma_{xy}$ is the sample cross correlation of **x** and **y** after removing their means. The items $C_1$, $C_2$, and $C_3$ are small positive constants that stabilize each term, so that near-zero sample means, variances, or correlations do not lead to numerical instability. The entire SSIM metric between the original and the reference image is calculated by averaging the local SSIM all over the image.

As mentioned before, we are to check the performance of the video coder with different distortion functions. We utilized the idea that MSE is not good enough to be used as quality metric and test the performance for various criterions. The main idea is to use SSIM to search for the best matching block in motion estimation of the video. Although the application of SSIM shows better performance than MSE and SAD (better block matching and thus better compression) but we go further to improve its performance. We explored the usage of CW-SSIM and VIF. A brief overview of both is given below.

### D. Complex Wavelet SSIM

As we saw above, SSIM measures the quality by comparing the structures in the image. As we know, small geometric distortions are not structural, so a new wavelet domain version of SSIM was introduced to overcome these artifacts [15]. CW-SSIM is usually formulated as [16]:

$$\tilde{S}(c_x, c_y) = \tilde{m}(c_x, c_y).\tilde{p}(c_x, c_y) = \frac{2\sum_{i=1}^{N}|c_{x,i}||c_{y,i}| + K}{\sum_{i=1}^{N}|c_{x,i}|^2 + \sum_{i=1}^{N}|c_{y,i}|^2 + K} \cdot \frac{2|\sum_{i=1}^{N}c_{x,i}.c^*_{y,i}| + K}{2\sum_{i=1}^{N}|c_{x,i}.c^*_{y,i}| + K} \quad (5)$$

Where in the previous equation, in the complex wavelet domain, $c_x = \{c_{x,i} | i=1,2,...,N\}$ and $c_y = \{c_{y,i} | i=1,2,...,N\}$ are respectively two sets of wavelet coefficients from the same spatial location in the same wavelet subbands of the two images being compared. $K$ is a small positive constant used for stabilizing. $\tilde{m}(c_x, c_y)$ is the SSIM index applied to the magnitude of the coefficients. $\tilde{p}(c_x, c_y)$ is calculated by monitoring the difference between phases of $c_x \& c_y$. First, CW-SSIM is calculated for each subband of the wavelet decomposition and then average of these values yields an overall CW-SSIM metric for the entire of the image. More details are available in [18].

*E. Visual Information Fidelity*

VIF quantifies the similarity of two images using a communication framework. It attempts to relate the signal fidelity to the amount of the information that is shared between the two signals (original and noisy version or distorted version). This shared information is quantified using the concept of mutual information which is widely used in information theory. Suppose that we are to compare the quality of two signals (images), a reference and a noisy version (in the current motion estimation application one signal is a block of pixels from reference frame and second is a block of pixels from next frame). Let us denote the reference signal by *C* and the distorted one by *D*. *E* is the perceived version of the source signal (*C*) by the neurons of the HVS (human visual system) and *F* is perceived version of *D*. We can write the following equations:

$$d = gc + v \quad , \quad e = c + n \quad , \quad f = d + n \qquad (6)$$

In these equations, *c,* and *d* are random vectors extracted from the same location of the same wavelet subband in the reference and distorted images, respectively. *g* is a scalar deterministic gain factor and *v* is an independent additive zero-mean white Gaussian noise. This model is a general model but has good performance almost everywhere. See [15] and [16] for more details. In receiver, the *n* is used to model the visual distortion as a stationary, zero-mean, additive white Gaussian noise process in the wavelet transform domain. The reference image is modeled by a wavelet domain Gaussian scale mixture (GSM), which is a good model for natural images [22]. Then *c* can be modeled as $c = \sqrt{z}\,u$ where *u* is a zero-mean Gaussian vector and $\sqrt{z}$ is an independent scalar random variable. According to [15] the VIF is computed by:

$$VIF = \frac{I(C;F|z)}{I(C;E|z)} = \frac{\sum_{i=1}^{N} I(c_i; f_i | z_i)}{\sum_{i=1}^{N} I(c_i; e_i | z_i)} \qquad (7)$$

where *i* is the index of local coefficients patches, with all subbands include. More information about VIF can be found in [15], [16].

### III. DATA

The video frames we used as the dataset are the standard Foreman video sequences. The format of the frames is standard CIF format. Figure 2 shows some sample frames used in the simulation. We have used the first frame as the reference frame and the second frame as the target frame in the simulation procedure.

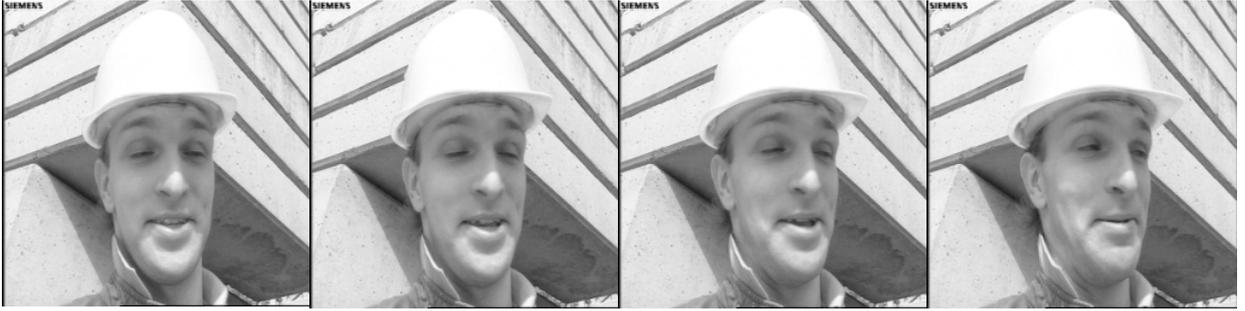

Figure 2. Sample frames used for simulations

### IV. METHOD

We modified the conventional motion estimation method with the usage of SSIM (and also CW-SSIM and VIF) as the similarity criterion for block matching. Macro blocks are selected with a side equal to 16 pixels (16×16 blocks). The comparison between our method and the conventional method is done in the full search approach i. e. the search region is a 48×48 block from the target frame. The best matched block in the search area is a 16×16 block with the biggest similarity to the selected 16×16 block from the reference frame. Biggest similarity is interpreted as biggest SSIM or biggest CW-SSIM or VIF. After that best match block is found, motion vectors are calculated and motion compensation is done using the motion vectors. The simulation results and trade-offs will be explained in the following section.

### V. RESULTS

Motion Compensation for Foreman frames with different quality criterions is simulated using a 2.2 GHz P.4 dual core processor. Table 1 shows the results of the simulation. As we can see from the table, VIF shows the best performance of compression. The constructed frame using motion vectors obtained by VIF is the most similar to the target frame. This similarity is computed by means of MSE, SSIM, VIF and also bitplane error. Bitplane error is the average Hamming distance between the reference frame and the target frame when each of them is considered as a sequence of binary bits (Each pixel is a 8 bit binary and the sequence can be obtained by putting these pixel values in serial order). The trade-off between the computational complexity and quality of the compressed video can be observed in the table. Although the usage of VIF shows better performance but its computational complexity is bigger than the other methods. It seems that SSIM is the knee point of the trade-off. The performance is acceptable while the complexity is not much bigger that MSE. The usage of VIF or CW-SSIM would be helpful in situations that the complexity is not of much concern such as in distributed video coding with motion estimation in the decoder [26]. Figure 3 shows the Foreman reference frame and target frame obtained by various metrics. Figure 3.a is the target frame. Figure 3.b, 3.d, 3.e, 3.f, 3.g are motion compensation output frames using SAD, MSE, SSIM, CW-SSIM, VIF quality criterions respectively. Figure 3.c is the constructed target frame with SAD distortion measure and a block of size 8 (instead of 16; we put this figure to sketch the poor performance of SAD). As we can see from the results the usage of quality assessment of the images in the motion compensation can help us to manage the video compression in case of required computational complexity and coding performance. Another interesting point is that the usage of SSIM, CW-SSIM and VIF more than reducing the average bitplane error, would cause in a remarkable decrease in the error of the most significant bits between the reference and target frames. Table 2 explains this fact. The left sided bits of the binary representation of the pixels have less error after the compression when we use the HVS based quality criterions in the motion estimation. The key point here is the fact that in many lossless and lossy compression techniques that only deal with the bit sequence (and the nature of the signal is not such important), more precise important bitplanes results in better coding. This is important because in almost all video coders and standards there is bit coding procedure (source coding stage) more than the motion compensation stage.

TABLE I. COMPARISON BETWEEN TARGET FRAME AND CONSTRUCTED FRAME USING MOTION COMPENSATION

| Quality | Comparison between target frame and constructed frame using motion compensation | Simulation Time (Sec.) |
|---|---|---|
| SAD | MSE=3.55, SSIM=0.55, VIF=0.09, dist=0.361 | 3 |
| MSE | MSE=3.33, SSIM=0.57, VIF=0.10, dist=0.350 | 3 |
| SSIM | MSE=13.16, SSIM=0.82, VIF=0.50, dist=0.256 | 7 |
| CWSSIM | MSE=10.42, SSIM=0.88, VIF=0.58, dist=0.249 | 16 |
| VIF | MSE=7.11, SSIM=0.91, VIF=0.63, dist=0.241 | 17 |

TABLE II. COMPARISON BETWEEN TARGET FRAME AND CONSTRUCTED FRAME USING MOTION COMPENSATION, BITPLANE HAMMING DISTANCES

| Quality | Comparison between target frame and constructed frame using motion compensation, bitplane Hamming distances |
|---|---|
| SAD | [0.164, 0.203, 0.322, 0.387, 0.428, 0.453, 0.366, 0.471] |
| MSE | [0.148, 0.189, 0.304, 0.364, 0.411, 0.448, 0.467, 0.473] |
| SSIM | [0.027, 0.042, 0.125, 0.203, 0.299, 0.413, 0.465, 0.475] |
| CWSSIM | [0.023, 0.040, 0.111, 0.195, 0.281, 0.401, 0.467, 0.478] |
| VIF | [0.019, 0.037, 0.095, 0.182, 0.273, 0.392, 0.463, 0.471] |

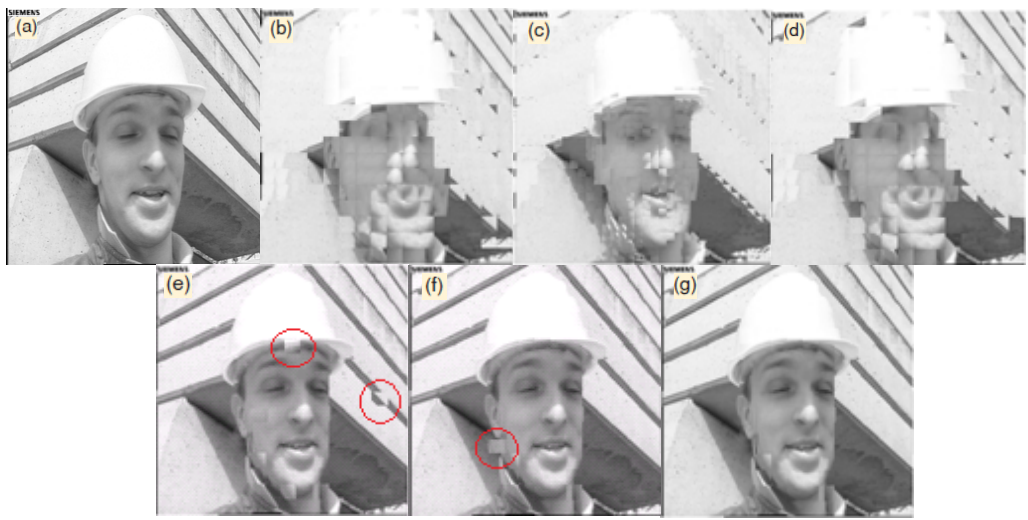

Figure 3. Comparison of the target frame and reconstructed frame using motion estimation: a) target frame b) construction using SAD c) SAD (8×8 macro block) d) MSE e) SSIM f) CW-SSIM g) VIF

## VI. CONCLUSION AND FUTURE WORKS

In this paper, we first review the recent advances in the video coding and specially motion compensation procedure then we state a brief overview about the image quality metrics. After that, we explain our method that is the usage of the HVS based metrics in the video compression and more precisely in motion compensation. We examine the legitimacy of our approach through the simulation. Results reviled this fact that usage of these metrics can make a remarkable improvement in the compression performance, especially on the motion compensation and source coding stages.

## VII. REFERENCES

[1] A. M. Tekalp, Digital Video Processing, Prentice Hall, Signal Processing Series, 1995.

[2] Introduction to Motion Picture Coding and the CCITT Algorithm, Inmos Limited, February 12, 1990.

[3] G. Ashraf and M. N. Chong, "Performance analysis of H.261 and H.263 video coding algorithms," IEEE Proceedings of International Symposium on Consumer Electronics, ISCE 97, Singapore, Dec. 1997.

[4] Sh. Uramoto, A. Takabatake, M. Suzuki, H. Sakurai and M. Yoshimoto, "A half-pel precision motion estimation processor for NTSC-resolution video," IEEE Conference on Custom Integrated Circuits, 1993.

[5] T. Toivonen and J. Heikkila, "Efficient method for half pixel block motion estimationusing block differentials," VLBV International Workshop, Sep. 2003, Madrid, Spain.

[6] Y. G. Lee, J. H. Lee and J. B. Ra, "Fast half pixel motion estimation based on directional search and a linear model," Proceedings of SPIE, IEEE Visual Communications and Image Processing, vol. 5150, pp. 1513-1520, 2003.

[7] T. Ebrahimi and C. Horne, "MPEG4 natural video coding, an overview," *IEEE Signal Processing: Image Communication*, vol. 15, pp. 365-385, 2000.

[8] B. Girod, E. Steinbach and N. Farber, "Comparison of H.263 and H.261 video compression standards," Standards and Common Interfaces for Video Information Systems, SPIE, vol. CR60, Oct. 1995, Philadelphia, USA.

[9] I. E. G. Richardson, *H.264 and MPEG-4 Video Compression*, England, John Wiley and Sons, 2003.

[10] K. R. Namuduri and A. Ji, "Computation and performance trade-offs in motion estimation," International Conference on Information Technology: Coding and Computing, ITCC 01, April 2001.

[11] T. Y. Kuo, J. Chalidabhongse and C. C. Jay Kuo, "Fast motion vector search for overlapped block motion compensation (OBMC)," Conference on Signals, Systems and Computers, ACSSC, vol. 2, pp. 948-952, 1996.

[12] E. Chan and S. Panchanathan, "Review of block matching based motion estimation algorithms for video compression," IEEE CCECE/CCGEI 1993.

[13] J. Y. Tham, S. Ranganath, M. Ranganath and A. Ali Kasim, "A novel unrestricted center biased diamond search algorithm for block motion estimation," *IEEE Transactions on Circuits and Systems for Video Technology*, vol. 8, No. 4, Aug. 1998.

[14] H. Jia and L. Zhang, "A new cross diamond search algorithm for block motion estimation," IEEE ICASSP 2004.

[15] H. R. Sheikh and A. C. Bovic, "Image information and visual quality," *IEEE Transactions on Image Processing*, Vol. 15, No. 2, Feb. 2006.

[16] Z. Wang and A. C. Bovik, "Mean squared Error: love it or leave it?," *IEEE Signal Processing Magazine,* Jan. 2009.

[17] Z. Wang, A. C. Bovik, H. R. Sheikh, and E. P. Simoncelli, "Image quality assessment: from error visibility to structural similarity," *IEEE Trans. Image Processing*, vol. 13, pp. 600–612, Apr. 2004.

[18] Z. Wang and E.P. Simoncelli, "Translation insensitive image similarity in complex wavelet domain," in *Proc. IEEE Int. Conf. Acoustics, Speech, Signal Processing*, Mar. 2005, pp. 573–576.

[19] Z. Wang, H. R. Sheikh and A. C. Bovik, *Objective Video Quality Assessment*, 'The Handbook of Video Databases: Design and Applications', CRC Press, Florida, USA, p.1041-1078, 2003.

[20] Z. Wang, L. Lu, A. C. Bovik, "Video quality assessment based on structural distortion measurement," *IEEE Signal Processing Magazine*, vol. 19, pp. 121-132, 2004.

[21] C. Cheung and L. Po, "Adjustable partial distortion search algorithm for fast block motion estimation," *IEEE Trans. on Circuits and Systems for Video Technology*, vol. 13, no. 1, Jan. 2003.

[22] J. Portilla, V. Strela, M. Wainwright, and E.P. Simoncelli, "Image denoising using scale mixtures of Gaussians in wavelet domain," *IEEE Trans. Image Processing*, vol. 12, no. 11, pp. 1338–1351, Nov. 2003.

[23] H. Zhang, Xi. Tian and Y. Chen, "A Video Structural Similarity Quality Metric Based on a Joint Spatial-Temporal Visual Attention Model", *Journal of Zhejiang Univ.*, Vol. 12, pp. 1696-1704, Springer, 2009.

[24] K. Seshadrinathan and A. C. Bovik, "A Structural Similarity Metric for Video Based on Motion Models", *IEEE Int. Conference on Acoustics, Speech and Signal Processing*, pp. 869-872, ICASSP 2007.

[25] A. Ortega and K. Ramchandran, "Rate-distortion methods for image and video compression," *IEEE Signal Processing Magezine,* pp. 23-50, Nov. 1998.

[26] J.Ascenso,C. Brites,and F. Pereira,"Improving Frame Interpolation with Spatial Motion Smoothing for Pixel Domain Distributed Video Coding," EURASIP Conference on Speech and Image Processing, Multimedia Communications and Services, Smolenice, Slovak Republic, June 2005.